\documentclass[conference]{IEEEtran}
\IEEEoverridecommandlockouts

\usepackage{cite}
\usepackage{amsmath,amssymb,amsfonts}
\usepackage{algorithmic}
\usepackage{graphicx}
\usepackage{textcomp}
\usepackage{xcolor}

\usepackage[caption = false]{subfig}
\usepackage{mathdots}
\usepackage{subfiles} 
\usepackage[export]{adjustbox}

\def\ie{{\textit{i.e.}}}
\def\eg{{\textit{e.g.}}}
\def\etal{{\textit{et al.}}}

\def\factorized{{Ball{\'{e}} \etal{} (2016)}}
\def\hyperprior{{Ball{\'{e}} \etal{} (2018)}}
\def\minnenjoint{{Minnen \etal{} (2018)}}
\def\cheng{{Cheng \etal{} (2020)}}
\def\nlaic{{Chen \etal{} (2021)}}

\def\zhihao{}

\begin{document}

\title{
Opening the Black Box of Learned Image Coders
}

\author{
\IEEEauthorblockN{Zhihao Duan}
\IEEEauthorblockA{
\textit{Purdue University}\\
West Lafayette, Indiana, U.S. \\
duan90@purdue.edu}
\and
\IEEEauthorblockN{Ming Lu}
\IEEEauthorblockA{
\textit{Nanjing University}\\
Nanjing, China \\
luming@smail.nju.edu.cn}
\and
\IEEEauthorblockN{Zhan Ma}
\IEEEauthorblockA{
\textit{Nanjing University}\\
Nanjing, China \\
mazhan@nju.edu.cn}
\and
\IEEEauthorblockN{Fengqing Zhu}
\IEEEauthorblockA{
\textit{Purdue University}\\
West Lafayette, Indiana, U.S. \\
zhu0@purdue.edu}
}


\maketitle

\begin{abstract}
End-to-end learned lossy image coders (LICs), as opposed to hand-crafted image codecs, have shown increasing superiority in terms of the rate-distortion performance. However, they are mainly treated as black-box systems and their interpretability is not well studied. In this paper, we show that LICs learn a set of basis functions to transform input image for its compact representation in the latent space, as analogous to the orthogonal transforms used in image coding standards. Our analysis provides insights to help understand how learned image coders work and could benefit future design and development.
\end{abstract}


\section{Introduction} \label{sec:intro}

\begin{figure}[t]
    \centering
    \subfloat[Linear orthogonal transformation]{
        \includegraphics[width=0.96\linewidth]{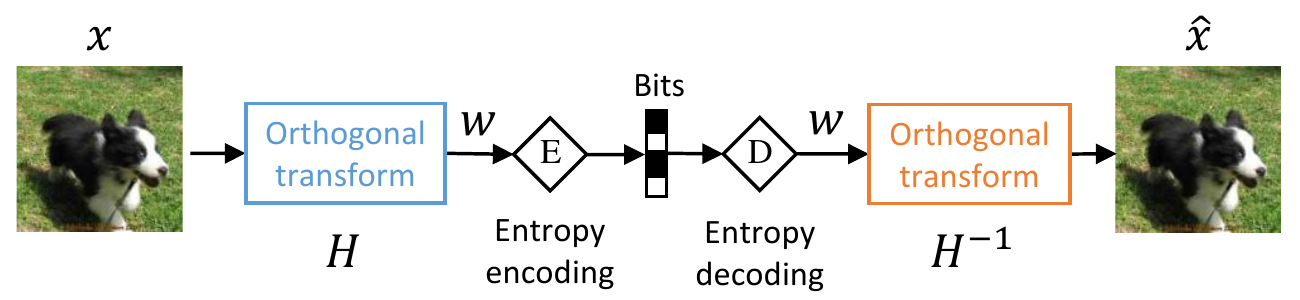}
        \label{fig:intro_linear}
    }
    \hfill
    \subfloat[Learned non-linear transformation]{
        \includegraphics[width=0.96\linewidth]{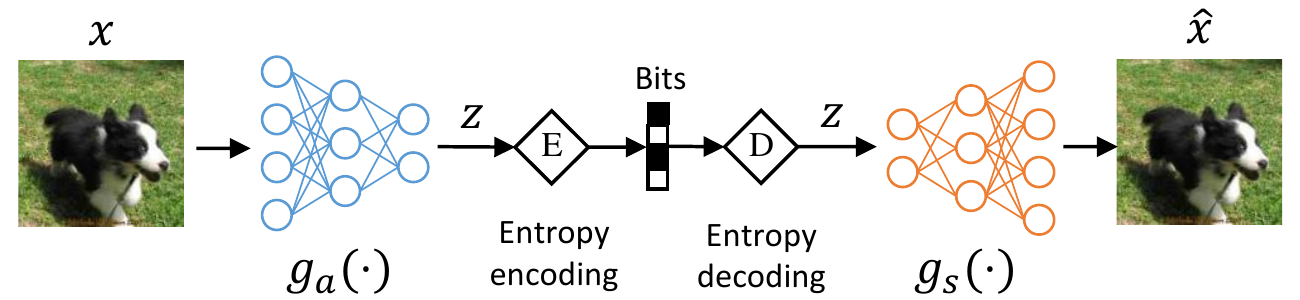}
        \label{fig:intro_learned}
    }
    \hfill
    \subfloat[``Basis functions" learned in LICs]{
        \includegraphics[width=0.96\linewidth]{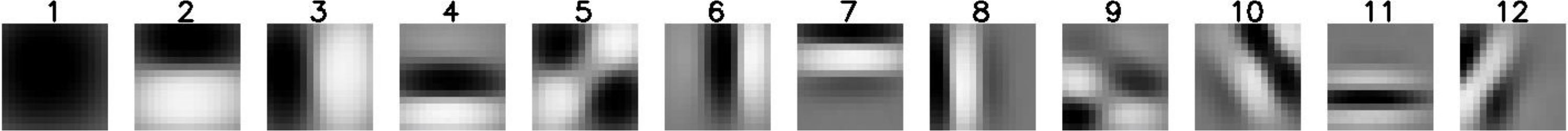}
        \label{fig:intro_basis}
    }
    \hfill
    \caption{\textbf{Typical frameworks of traditional (a) and learning-based (b) image coding.} Entropy models are omitted for simplicity.
    This paper shows that LICs learn to transform images according to a set of ``basis functions", which is similar to the linear orthogonal basis in traditional image codecs.
    }
    \label{fig:intro}
\end{figure}

Image pixels are highly correlated. The core idea of \textit{transform coding} is to transform image pixels into a compact representation in the sense that the coefficients are ideally decorrelated and the total entropy is concentrated on a few of them, with which we can code the compact representation instead of the pixel values.
For conventional image coding (Fig.~\ref{fig:intro_linear}), the transformation module typically  relies on orthogonal linear mapping functions, such as the discrete cosine transform (DCT) in JPEG~\cite{wallace1992jpeg} and DCT-alike integer transform in HEVC/VVC intra coding~\cite{lainema2012hevc_intra}.
Learned image coders (LICs), however, use non-linear neural networks (Fig.~\ref{fig:intro_learned}) to fulfill such transformations, and the network parameters are learned to optimize the rate-distortion loss  on training  images~\cite{balle2016end2end}.

Such learning-based approach has been rapidly improved over recent years~\cite{theis2017lossy, minnen2020channelwise, cheng2020cvpr,chen2021nlaic}, being on par with the state-of-the-art hand-crafted codec (\ie, VVC intra~\cite{pfaff2021vvc_intra}).
Like most deep learning-based systems, LICs are less interpretable compared with hand-crafted algorithms.
In comparison to \textit{entropy models} that characterize the data distribution of latent features in LICs~\cite{minnen2018joint, minnen2020channelwise}, the non-linear \textit{transformation} remains poorly understood.
Unlike in traditional coders, where the linear transformation can be fully described by a set of orthogonal basis vectors, the LIC transformation is difficult to analyze due to the use of deep layers and layer-wise non-linear activation, and thus are commonly treated as a black-box module fully relying on the data driven training to determine the parameters.
In this paper, we take a step towards opening the black-box of LICs by characterizing the non-linear transformation from a basis decomposition perspective.




\zhihao{We begin by noticing that the compressed coefficients ($\mathbf{z}$ in Fig.~\ref{fig:intro_learned}) extracted by LICs could reflect the functionality of LIC transformations.}
We decode each compressed coefficient individually, and we observe that each compressed coefficient decodes to a unique pattern (Fig.~\ref{fig:intro_basis}) that is visually similar to the basis functions of linear orthogonal transformations.
We thus extend the definition of ``basis functions", which is originally defined for linear transformations, to the non-linear transformations in LICs.
Extensive experiments show that similar basis functions consistently occur in various LICs,
being independent to network architectures, bit rates, and reconstruction loss functions.
Motivated by the similarity between the framework of linear transform coding (Fig.~\ref{fig:intro_linear}) and LICs (Fig.~\ref{fig:intro_learned}), we empirically conclude that LICs learn a non-linear counterpart of orthogonal transform coding,
which coincides with the hand-crafted design in traditional codecs.

Our contributions can be summarized as follows. We 
define the basis functions of learned image coders (LICs)
to analyze LICs from the perspective of orthogonal transform coding.
We conduct experiments on a wide range of LIC designs including different architectures, entropy models, distortion metrics, and bit rates.  
Our results and analysis help to understand how LICs work, as well as provide insights to improve future LICs design.

\section{Background and Related Works} \label{sec:related}

\subsection{Linear Transform Coding}
\label{sec:linear_transform_coding}
In linear transform coding, an image patch (viewed as a column vector) $\mathbf{x} \in \mathbb{R}^d$ is projected onto the transform space by an orthogonal (or orthonormal) matrix $H \in \mathbb{R}^{d\times d}$:
\begin{equation}
    \label{eq:linear_transform}
    \mathbf{w} = H \mathbf{x},
\end{equation}
where $\mathbf{w}$ is the transform coefficients. The rows of $H$ is known as the transform basis vectors, and the transform coefficients indicate the contribution of each basis to represent the original image.
\zhihao{For example, in JPEG, $H$ would be the discrete cosine transform (DCT) matrix, and its rows are known as the DCT basis functions.}
An orthogonal linear transform \zhihao{can be fully described by} the set of basis functions.

\subsection{Learned Image Coding}
Typically, learned image coders use convolutional neural networks (CNNs) to construct an analysis transform $g_a(\cdot)$ (or \textit{encoder}), a synthesis transform $g_s(\cdot)$ (or \textit{decoder}), and an entropy model $p_Z(\cdot)$ to compress images.
Given an image $\mathbf{x}$, the LIC transformations can be formulated as follows:
\begin{equation}
    \label{eq:lic_transform}
    \begin{aligned}
    \mathbf{z} &= g_a(\mathbf{x})
    \\
    \hat{\mathbf{x}} &= g_s(\mathbf{z}),
    \end{aligned}
\end{equation}
where $\mathbf{z}$ is a three-dimensional (channel, height, and width) array, which we refer to as the \textit{compressed representation}.
Note that we omit quantization and entropy models for simplicity.
In LICs, the network parameters are learned from data by minimizing the empirical rate-distortion loss function~\cite{balle2016end2end}.

This framework of LIC has different interpretations.
\zhihao{
Ball{\'{e}} \etal{} have shown that such framework can be viewed as variational autoencoders (VAEs)~\cite{balle2016end2end, balle18hyperprior}.
}
Alternatively, it can also be viewed as a vector quantizer powered by learned, non-linear transformations~\cite{balle2021nonlinear}.
However, important components including $g_a(\cdot)$ and $g_s(\cdot)$ are mostly treated as black-boxes in previous works, lacking explanations or insights of their specific functions. 
In this paper, we aim to open the black box of LIC transformations using the methods proposed in Section~\ref{sec:main}.


\section{Interpreting Learned Image Coders} \label{sec:main}

In this section, we describe how we visualize and interpret learned image coders (LICs).
We begin by analyzing
each individual coefficient
in the compressed representation.
Later, we define the basis functions and use them to interpret LICs.

\begin{figure}[t]
    \centering
    \subfloat[Illustration of spatial decomposition.]{
        \includegraphics[width=0.92\linewidth]{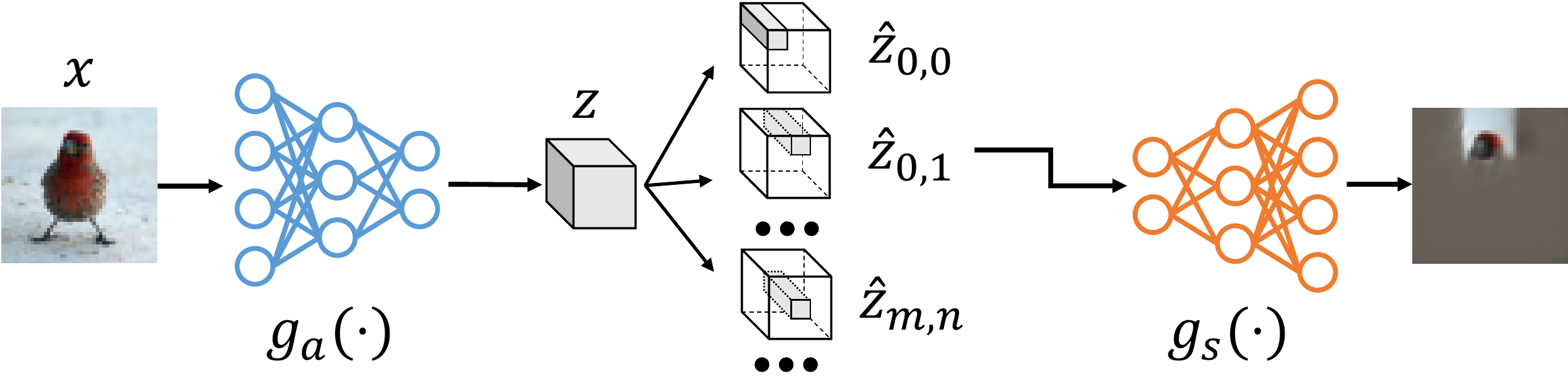}
        \label{fig:method_decompose_spatial}
    }
    \hfill
    \subfloat[Illustration of channel-wise decomposition.]{
        \includegraphics[width=0.92\linewidth]{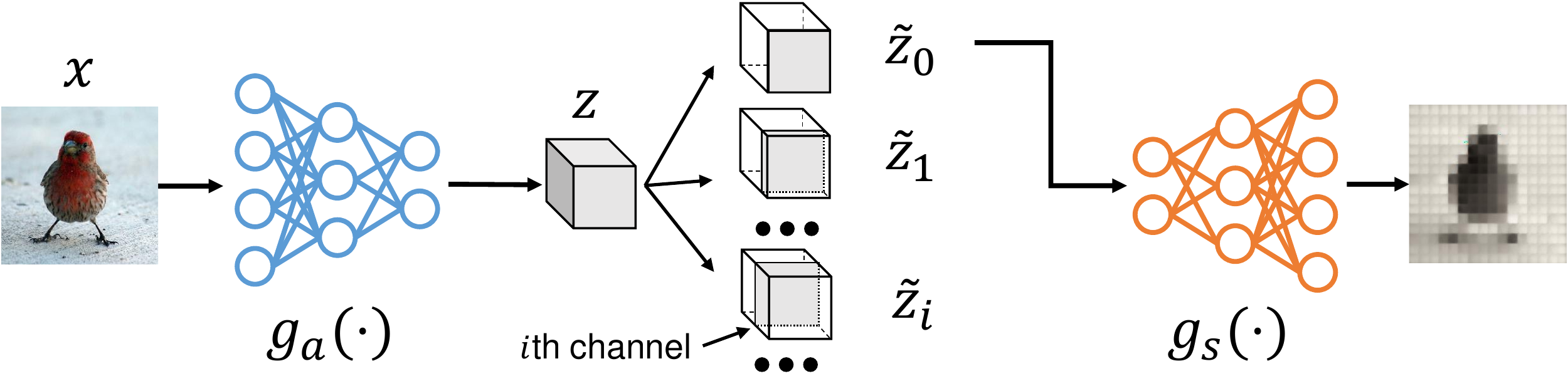}
        \label{fig:method_decompose_channel}
    }
    \hfill
    \subfloat[Results (spatial).]{
        \includegraphics[width=0.46\linewidth]{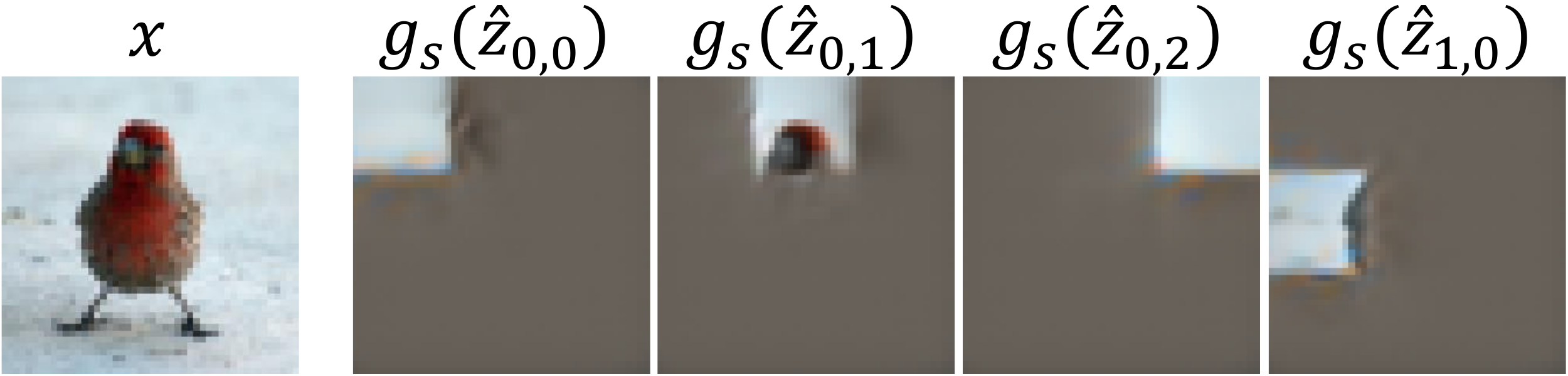}
        \label{fig:method_decompose_spatial_results}
    }
    \subfloat[Results (channel-wise).]{
        \includegraphics[width=0.46\linewidth]{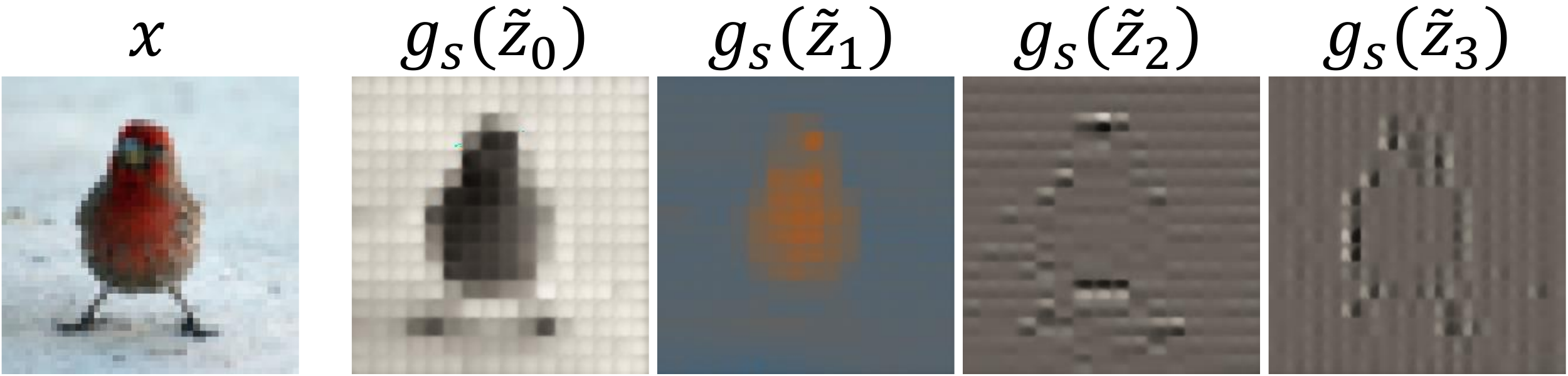}
        \label{fig:method_decompose_channel_results}
    }
    \caption{\textbf{Compressed representation decomposition.}
    In (a) and (b), we show how we decompose the compressed representation into subsets.
    In (c) and (d), we show the reconstructions using the decomposed compressed coefficients, where we scale the input image so that the compressed representation has a spatial resolution of $3\times 3$ in (c) and $16\times 16$ in (d).
    In (d), we sort the channels by their bit rates in descending order. The LIC model is from Minnen \etal~\cite{minnen2018joint}.
    Best viewed by zooming in.
    }
    \label{fig:method_decompose}
\end{figure}

\subsection{
Decomposing the Compressed Representation
} \label{sec:method_decompose}

In traditional linear coders,
the compressed coefficients (\eg, DCT coefficients) of an image are fully interpretable, as each coefficient represents a specific frequency component in the original image.
Motivated by this, we ask the question: can we also interpret the compressed representation of LICs?
To answer this question, we propose a heuristic solution: we decompose the compressed representation of an image into different subsets and decode each subset separately. Then, the decoded ``image" using each coefficient subset reflects the information stored in that subset.
We describe our method more formally below.


\textbf{Spatial decomposition.}
Recall that in LICs, the compressed representation $\mathbf{z}$ of an image is a three-dimensional array (channel, height, and width).
Given $\mathbf{z}$, we can decompose it along the height and width dimension:
\begin{equation}
    \mathbf{z} = \sum_{m,n} \hat{\mathbf{z}}_{m,n},
\end{equation}
where each $\hat{\mathbf{z}}_{m,n}$ is defined as an all-zero array except that we assign $\hat{\mathbf{z}}_{m,n}[:,m,n] \triangleq \mathbf{z}[:,m,n]$. That is, $\hat{\mathbf{z}}_{m,n}$ contains only the coefficients of $\mathbf{z}$ at spatial index $m,n$.
Then, we decode each $\hat{\mathbf{z}}_{m,n}$ using the synthesis transform $g_s(\cdot)$, and we hypothesize that the decoded ``image", $g_s(\hat{\mathbf{z}}_{m,n})$, indicates the image component stored in $\hat{\mathbf{z}}_{m,n}$.
The complete procedure is illustrated in Fig.~\ref{fig:method_decompose_spatial}.

\textbf{Channel-wise decomposition.}
We can do the same decomposition channel-wisely, as shown in Fig.~\ref{fig:method_decompose_channel}. Formally,
\begin{equation}
    \mathbf{z} = \sum_{i} \tilde{\mathbf{z}}_{i},
\end{equation}
where each $\tilde{\mathbf{z}}_{i}$ is defined as an all-zero array except that we assign $\tilde{\mathbf{z}}_{i}[i,:,:] \triangleq \mathbf{z}[i,:,:]$. By decoding $\tilde{\mathbf{z}}_{i}$, we hypothesize that $g_s(\tilde{\mathbf{z}}_{i})$ visualize the image component stored in the $i$th channel of $\mathbf{z}$.

\textbf{Observations.}
We show example results for decomposition followed by reconstruction in Fig.~\ref{fig:method_decompose_spatial_results} and Fig.~\ref{fig:method_decompose_channel_results}.
Due to the limited space, we only show results for the Joint AR \& H model by Minnen \etal~\cite{minnen2018joint}, but in our experiments we observe similar results for all LICs we tested.
In Fig.~\ref{fig:method_decompose_spatial_results}, we first observe that each feature vector $\mathbf{z}[:,m,n]$ mostly contributes to only a patch of the reconstructed image. This indicates that the LIC transformations are highly localized, which is presumably due to the extensive use of convolutional layers in LICs.
In Fig.~\ref{fig:method_decompose_channel_results}, we observe that the channel-wise decoding share patterns with the original image. For example, $\hat{\mathbf{z}}_0$ captures the brightness and $\hat{\mathbf{z}}_1$ captures the color of $\mathbf{x}$. More interestingly, $\hat{\mathbf{z}}_2$ and $\hat{\mathbf{z}}_3$ respond to the vertical and horizontal edges in the original image, respectively.

By decomposing and decoding the compressed coefficients,
we find that each coefficient in the LIC latent space potentially has a human-interpretable meaning.
Along this direction, we give a more general method for interpreting LIC compressed coefficients, by which we can characterize the functionality of LIC transformations, in the next section.

\subsection{Basis Functions of LICs} \label{sec:method_basis}
To fully interpret the compressed representation $\mathbf{z}$, one could decompose all the coefficients into $\hat{\mathbf{z}}_{i,m,n}$ similarly as in the previous section.
However, if we assume convolutional networks to be block-wise shift-invariant~\cite{zhang2019making_cnn_shift_invariant}, we can avoid enumerating all spatial indices $m,n$.
To also avoid the dependency on specific images, we manually design the compressed representations instead of using the ones from real images.

\begin{figure}[t]
    \centering
    \subfloat[Basis function of LICs.]{
        \includegraphics[width=0.42\linewidth]{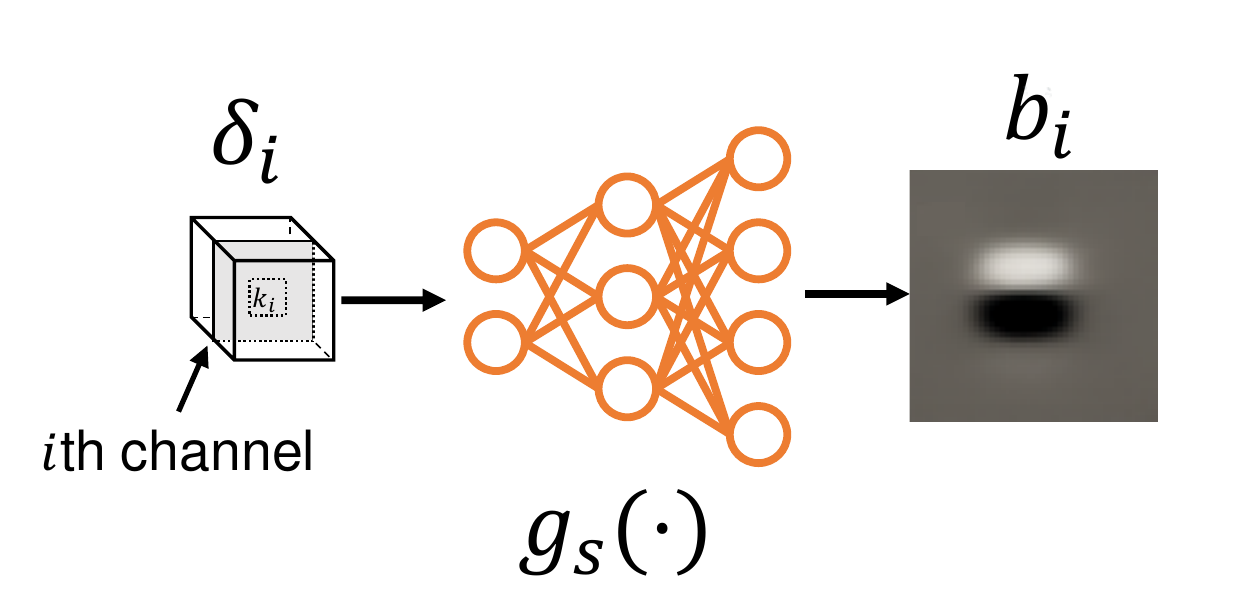}
        \label{fig:method_main_basis}
    }
    \subfloat[Computing $\text{MSE}_\text{channel}$.]{
        \includegraphics[width=0.5\linewidth]{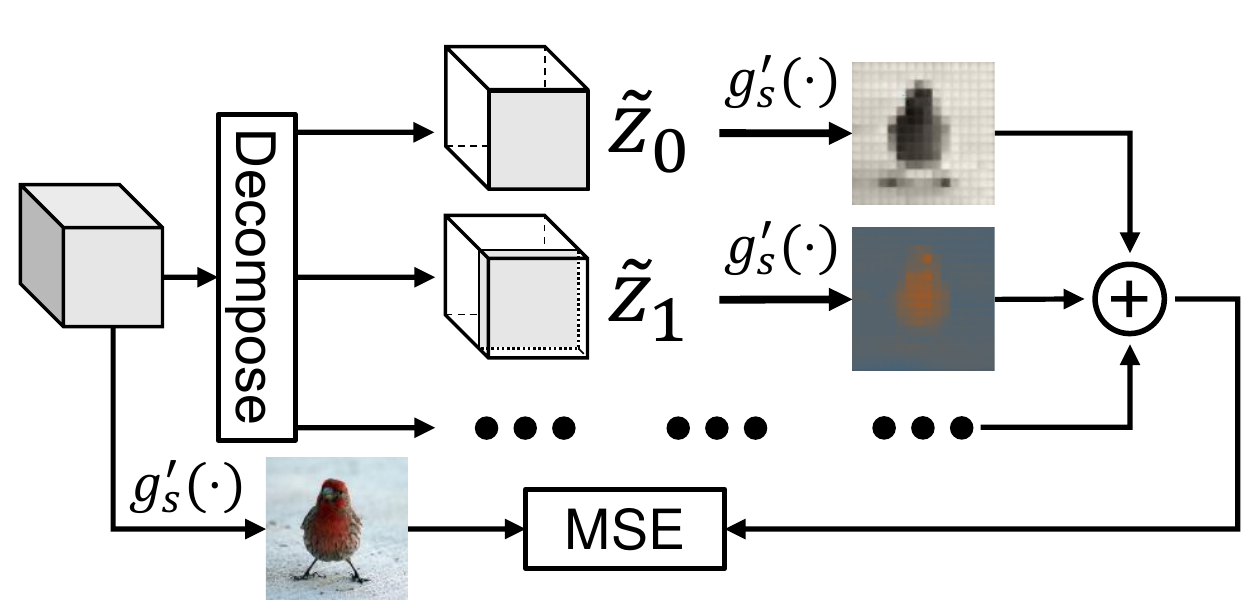}
        \label{fig:method_main_independency}
    }
    \hfill
    \caption{Illustration of (a) visualizing LIC basis functions, and (b) measuring $\text{MSE}_\text{channel}$ as defined in \eqref{eq:channel_spatial_independency}. The computation of $\text{MSE}_\text{spatial}$ is done similarly.
    }
    \label{fig:method_main}
\end{figure}

Specifically, we define our ``artificial" compressed representation, $\boldsymbol{\delta}_i$, as 
a three-dimensional (channel, height, and width) integer array, in which all elements of $\boldsymbol{\delta}_i$ are set to zero except the center element of the $i$th channel:
\begin{equation}
    \boldsymbol{\delta}_i[i,:,:] \triangleq \begin{bmatrix}
        \ddots &  & \vdots &  & \iddots \\ 
         & 0 & 0 & 0 & \\ 
        \dots & 0 & k_i & 0 & \dots\\ 
         & 0 & 0 & 0 & \\ 
         \iddots &  & \vdots &  & \ddots
    \end{bmatrix},
\end{equation}
where $k_i\in \mathbb{Z}$ is a real number that we can tune.
Then, we decode $\boldsymbol{\delta}_i$ by the synthesis transform $g_s(\cdot)$, and define the output to be the \textit{basis functions} (more details in Sec.~\ref{sec:method_independency}) of LICs:
\begin{equation}
    \mathbf{b}_i \triangleq g_s(\boldsymbol{\delta}_i), \ \forall i = 1,2,...
    \label{eq:lic_basis}
\end{equation}
This process is illustrated in Fig.~\ref{fig:method_main_basis}.
From a signal processing perspective, a CNN-based synthesis transform is a (non-linear) shift-invariant system, so each $\mathbf{b}_i$ can be viewed as the impulse response for each ``channel impulse", $\boldsymbol{\delta}_i$.

With the above definition, we argue that the ``channel impulse responses", $\mathbf{b}_i$, are non-linear counterparts of basis vectors, which are originally defined for linear transformations.
Recall that in linear coders, the basis vectors can be obtained by selecting the rows in the orthogonal transform matrix $H$, or equivalently, the columns of its inverse $H^{-1}$:
\begin{equation}
    \mathbf{b}_i^\text{linear} = H^{-1} \boldsymbol{\delta}_i^\text{linear}, \ \forall i = 1,2,...,
    \label{eq:linear_basis}
\end{equation}
where $\mathbf{b}_i^\text{linear}$ is the $i$th basis vector, and $\boldsymbol{\delta}_i^\text{linear}$ is a column-selecting vector, \ie, its elements are all zero except the $i$th element being $1$:
\begin{equation}
    \boldsymbol{\delta}_i^\text{linear} \triangleq [\dots, \ 0, \underset{i\text{th}}{1}, 0, \ \dots]^T.
\end{equation}
Comparing \eqref{eq:lic_basis} and \eqref{eq:linear_basis}, we can find correspondence between the decoders ($g_s(\cdot)$ and $H^{-1}$) as well as impulse inputs ($\boldsymbol{\delta}_i$ and $\boldsymbol{\delta}_i^\text{linear}$).
Then, our $\mathbf{b}_i$ can be interpreted as the non-linear counterparts of (linear) basis vectors $\mathbf{b}_i^\text{linear}$.
Similar to the channel-wise decomposition, these $\mathbf{b}_i$ provide intuition about the role of the $i$th channel in the compressed domain of LICs. In addition, they are more general and explicit, since $\mathbf{b}_i$ do not depend on any specific images.
In our experiments (Sec.~\ref{sec:results}), we show that $\mathbf{b}_i$ resemble linear orthogonal basis, and they can be interpreted as the basis functions of LICs.

There are several details of the above definition of LIC basis that worth noting.
First, the height and width dimensions of $\boldsymbol{\delta}_i$ can be chosen arbitrarily without affecting the output $g_s({\boldsymbol{\delta}}_i)$, since the convolution layers in $g_s(\cdot)$ zero-pad the input anyway. We set them to be $1\times 1$ in our experiments for simplicity.
Also, the impulse magnitude $k_i$, which control the amplitude of $g_s(\boldsymbol{\delta}_i)$, can also be set arbitrarily. In our experiments, we choose $k_i$ to be the largest value of the $i$th channel of the compressed representations of the Kodak~\cite{kodak} images.
Finally, as $g_a(\cdot)$ and $g_s(\cdot)$ (\ie, encoder and decoder) can be viewed as a conceptual inverse of each other, we assume that by using the $g_s(\cdot)$ alone, we can generalize our conclusions to both of them.

\begin{figure*}[ht]
    \centering
    \subfloat[\factorized{}~\cite{balle2016end2end}, MSE loss, 0.487 bpp. Trained on grayscale images.]{
        \includegraphics[width=0.98\linewidth]{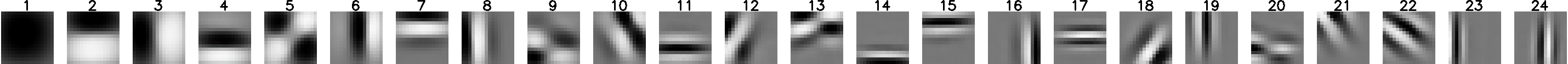}
         \label{fig:basis_gray}
    }
    \hfill \vspace{0.1cm}
    \subfloat[\factorized{}~\cite{balle2016end2end}, MSE loss, 0.647 bpp. Trained on color images.]{
        \includegraphics[width=0.98\linewidth]{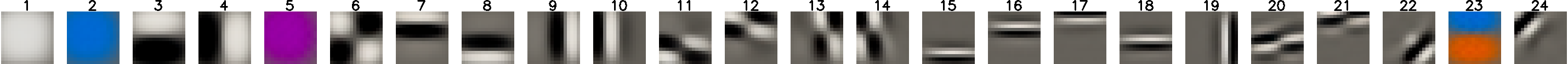}
         \label{fig:basis_factorized_mse}
    }
    \hfill \vspace{0.1cm}
    \subfloat[\hyperprior{}~\cite{balle18hyperprior}, MSE loss, 0.937 bpp. Trained on color images.]{
        \includegraphics[width=0.98\linewidth]{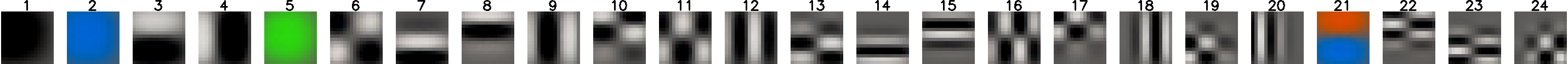}
         \label{fig:basis_hyper_mse}
    }
    \hfill \vspace{0.1cm}
    \subfloat[\minnenjoint{}~\cite{minnen2018joint}, MSE loss, 0.195 bpp. Trained on color images.]{
        \includegraphics[width=0.98\linewidth]{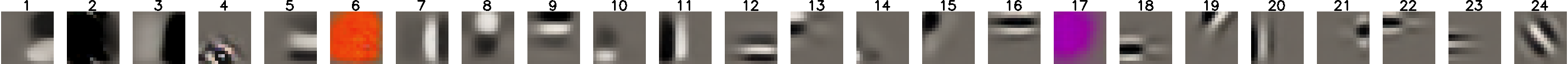}
         \label{fig:basis_hyper_msssim}
    }
    \hfill \vspace{0.1cm}
    \subfloat[\cheng{}~\cite{cheng2020cvpr}, MS-SSIM loss, 0.315 bpp. Trained on color images.]{
        \includegraphics[width=0.98\linewidth]{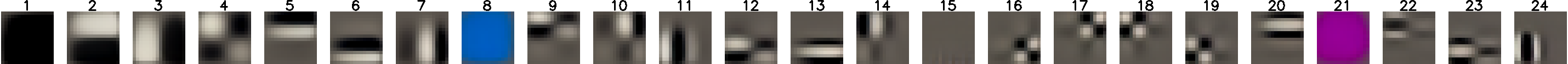}
         \label{fig:basis_cheng_msssim}
    }
    \hfill \vspace{0.1cm}
    \subfloat[\nlaic{}~\cite{chen2021nlaic}, MS-SSIM loss, 0.133 bpp. Trained on color images.]{
        \includegraphics[width=0.98\linewidth]{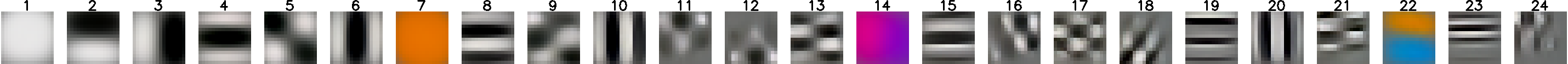}
         \label{fig:basis_nlaic_msssim}
    }
    \hfill
    \caption{\textbf{Channel basis of various LICs across different bit rates.} 
    Each sub-image (with resolution $16\times 16$) corresponds to one coefficient of the compressed representation.
    For each LIC, channels are sorted in decreasing order by their bit rates on the Kodak~\cite{kodak} image set, and the top-24 channels with the highest bit rates are shown. The channel index (after sorting) is labeled on top of each sub-image. Image brightness is scaled for better visualization.}
    \label{fig:basis}
\end{figure*}

\subsection{\zhihao{Separability} Hypothesis} \label{sec:method_independency}
Our decomposition (spatial and channel-wise in Sec.~\ref{sec:method_decompose}, and element-wise in Sec.~\ref{sec:method_basis}) of $\mathbf{z}$ implicitly assumes that the coefficients of $\mathbf{z}$ are \zhihao{``separable"}, in the sense that they can be decoded separately and then aggregated to form
the image which can be normally decoded to (\ie, $g_s(\mathbf{z})$).
In this section, we propose methods to validate our hypothesis of independence.
Notice that in our context, \textit{independence} does not imply independent random variables.
To verify that the compressed coefficients can be decomposed and decoded separately, we propose to measure the difference between
this separate decoding and the normal image reconstruction,
and if this difference is small, we can confirm our \zhihao{separability} hypothesis.
We propose to measure this difference in two directions, spatially and channel-wise, using the following quantities:
\begin{equation}
\begin{aligned}
    \text{MSE}_\text{spatial} &\triangleq \text{MSE}(g'_s(\mathbf{z}), \sum_{m,n} g'_s(\hat{\mathbf{z}}_{m,n})) \\
    \text{MSE}_\text{channel} &\triangleq \text{MSE}(g'_s(\mathbf{z}), \sum_i g'_s(\tilde{\mathbf{z}}_i)),
\end{aligned}
\label{eq:channel_spatial_independency}
\end{equation}
where $\text{MSE}(\cdot)$ denotes the mean square error function, $g'_s(\cdot) \triangleq g_s(\cdot) - g_s(\mathbf{0})$ is the synthesis transform without offset, and $\hat{\mathbf{z}}_{m,n}, \tilde{\mathbf{z}}_{i}$ are the decompositions defined in Sec.~\ref{sec:method_decompose}.

The above procedure is also illustrated in Fig.~\ref{fig:method_main_independency}.
It measures the difference between the normally decoded image, $g'_s(\mathbf{z})$, and the ones where $\mathbf{z}$ is decomposed, separately decoded, and then aggregated.
\zhihao{In the ideal situation where the compressed coefficients can be separately decoded (\eg, in the linear case)}, we would have both $\text{MSE}_\text{spatial}$ and $\text{MSE}_\text{channel}$ being 0.
So, a small value of MSE would support our \zhihao{separability} hypothesis,
and thus our definition of $\mathbf{b}_i$ in \eqref{eq:lic_basis} can be safely interpreted as the basis functions of LICs.

\section{Results and Discussion} \label{sec:results}




We first hypothesize that our \zhihao{separability} hypothesis is true and visualize the basis functions for various LICs in Sec.~\ref{sec:exp_basis}.
We then validate the hypothesis in Sec.~\ref{sec:exp_assumption}.
Finally, we discuss our results and future work in Sec.~\ref{sec:discussion}.

\subsection{Basis Functions of LICs} \label{sec:exp_basis}
We start with a simple case where a LIC with a factorized entropy model~\cite{balle2016end2end} is trained on grayscale images.
We plot the basis functions for this gray image coder in Fig.~\ref{fig:basis_gray}, where the channels are sorted by their bit rates on the Kodak image set (ranks are labeled on each basis).
Then, we do the same thing for various LICs for color images in Fig.~\ref{fig:basis_factorized_mse}-\ref{fig:basis_nlaic_msssim}.

Our first observation is that, in Fig.~\ref{fig:basis_gray}, the basis functions of a grayscale LIC is surprisingly similar to orthogonal transform basis, such as Walsh-Hadamard Transform~\cite{ahmed1975orthogonal} and orthogonal wavelets.
This motivates us to interpret the compressed representation as orthogonal transform coefficients.
When moving from grayscale images to color images, there are two further interesting observations: 1) such basis pattern retains across all cases, being invariant to model architectures, distortion metrics, and bit rates; 2) there emerges chroma components (\eg, the $7$th, $14$th, and $22$nd channel in Fig.~\ref{fig:basis_nlaic_msssim}), which are independent of luma components.



\subsection{Validation of Hypothesis} \label{sec:exp_assumption}
Recall that, in Sec.~\ref{sec:method_independency}, we need $\text{MSE}_\text{channel}$ and $\text{MSE}_\text{spatial}$ to be close to zero to safely interpret $\mathbf{b}_i$ as the transform basis functions.
To verify this, we measure them using various LICs on the Kodak image set and present the results in Table~\ref{table:linearity}, where image pixel value ranges from 0 to 1, and the LICs are sorted by their rate-distortion performance in ascending order (from top to bottom).
We average $\text{MSE}_\text{channel}$ over all images but compute $\text{MSE}_\text{spatial}$ only on the first image due to its high computational complexity.
\zhihao{We also show the corresponding standard deviation (std.) computed over all pixels.}
We can see that both of $\text{MSE}_\text{channel}$ and $\text{MSE}_\text{spatial}$ are less than $0.005$ \zhihao{and that the standard deviations are close to zero} in all cases, which supports our \zhihao{separability} hypothesis.

\begin{figure}[t]
    \centering
    \subfloat[$\sum_i g'_s(\tilde{\mathbf{z}}_i)$]{
        \includegraphics[width=0.3\linewidth]{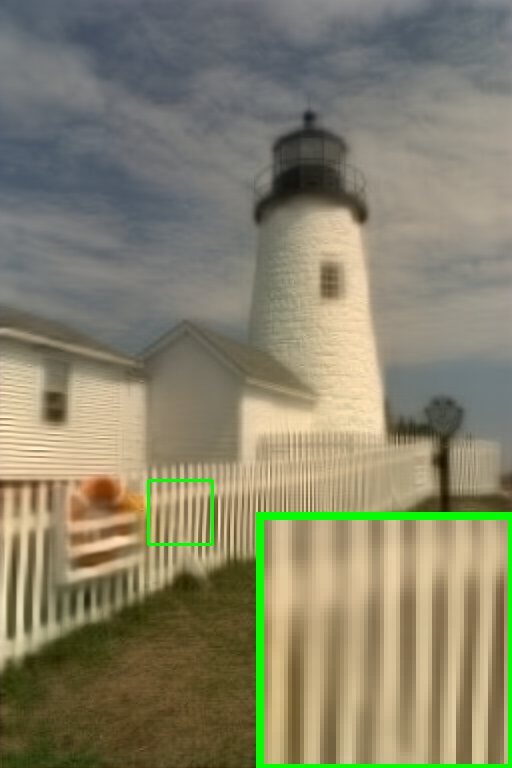}
    }
    \subfloat[$\sum_{m,n} g'_s(\hat{\mathbf{z}}_{m,n})$]{
        \includegraphics[width=0.3\linewidth]{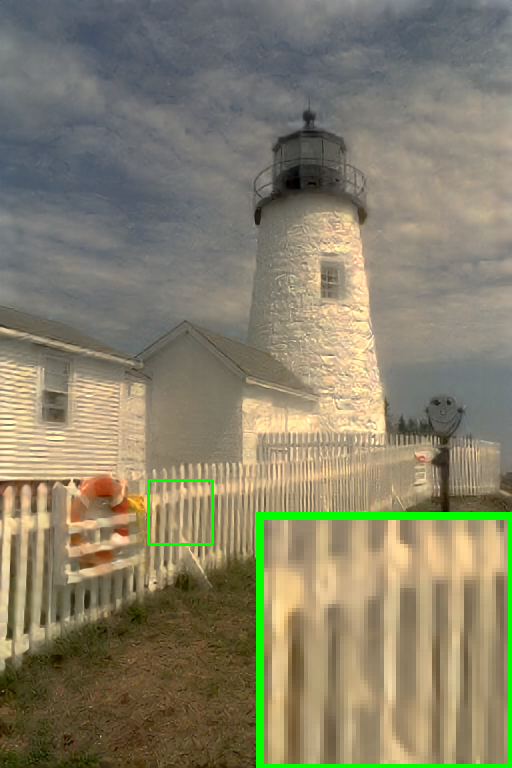}
    }
    \subfloat[$g'_s(\mathbf{z})$]{
        \includegraphics[width=0.3\linewidth]{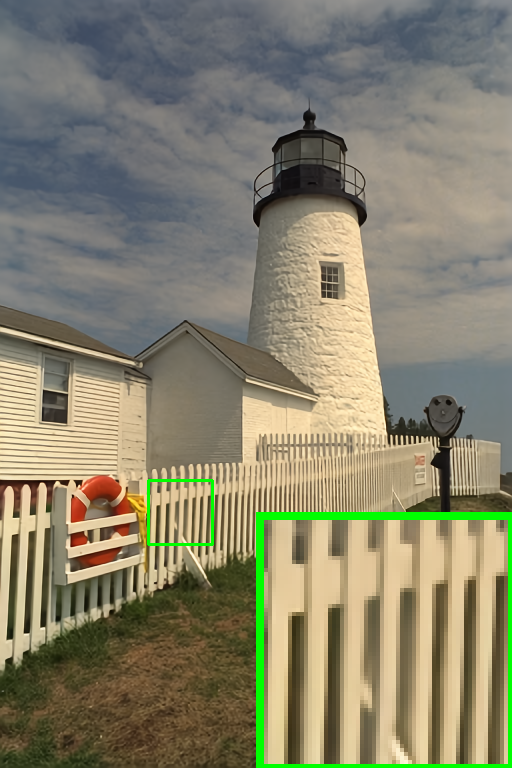}
    }
    \caption{Qualitative comparison for channel-wise and spatial independency. Terms are defined in Sec.~\ref{sec:method_independency}.
    In (a), we decompose $\mathbf{z}$ channel-wise, decode each subset, and aggregate the results. In (b), we do the same procedure but spatially. We show the the normal, joint decoding in (c).}
    \label{fig:linearity}
\end{figure}

We also show a qualitative example using \nlaic{} in Fig.~\ref{fig:linearity}.
We can visually observe that the channel-wise and spatial decomposition introduce blurring artifacts,
but both of them produce reasonable reconstructions of the original image.



\begin{table}[t]
\centering
\caption{Measurement of channel-wise and spatial \zhihao{separability}.
}
\begin{tabular}{l|c|c}
\hline
               & $\text{MSE}_\text{channel}$ (std.) & $\text{MSE}_\text{spatial}$ (std.) \\ \hline
\factorized{}  & 0.0026 (0.0084)                    & 0.0006 (0.0021)                    \\
\hyperprior{}  & 0.0026 (0.0104)                    & 0.0004 (0.0012)                    \\
\minnenjoint{} & 0.0022 (0.0084)                    & 0.0015 (0.0053)                    \\
\cheng{}       & 0.0037 (0.0124)                    & 0.0031 (0.0092)                    \\
\nlaic{}       & 0.0045 (0.0138)                    & 0.0046 (0.0112)                    \\ \hline
\end{tabular}
\label{table:linearity}
\end{table}

\subsection{Discussion}
\label{sec:discussion}
From Fig.~\ref{fig:basis}, we conclude that there are two internal mechanisms in the LIC analysis transform. It first performs an RGB to luma-chroma conversion, and then it performs a basis decomposition for each of the luma and chroma component, respectively (\eg, comparing the 2nd and 22nd channel in Fig.~\ref{fig:basis_nlaic_msssim}).
Interestingly, such basis decomposition closely resembles the orthogonal transformations that are widely adopted in image processing.
For example, the LIC basis of \cheng{} (Fig.~\ref{fig:basis_cheng_msssim}) is visually similar to Haar wavelets~\cite{porwik2004haar_wavelet}, while the ones of \nlaic{} (Fig.~\ref{fig:basis_nlaic_msssim}) are more like the basis of 2-D Walsh-Hadamard Transform.

We also notice the surprising similarity between this learned behavior and conventional hand-crafted codecs, such as the RGB-to-YCbCr conversion and DCT in JPEG.
In fact, the optimal linear transform that minimizes basis restriction error (\ie, only keep a subset of transform coefficients and discard the others) is the Karhunen–Loeve transform (KLT), which is known to be similar to DCT on images~\cite{ahmed1974dct}. We thus heuristically conclude that the LIC transforms can be viewed as a non-linear counterpart of KLT, and the LIC basis is the optimal basis computed on the training set. However, a rigorous proof is nontrivial, and should be pursued in future work.

The basis decomposition property of different LICs as well as linear coders also raises an interesting question: what are the key contributing factors that make LICs perform better?
We attribute this to two advances in LICs: the network architecture and learned entropy models.
It is well-known that the network architecture can largely impact a model's performance
in a wide range of image processing tasks~\cite{lim2017edsr, liang2021swinir} as well as in image compression~\cite{cheng2020cvpr, chen2021nlaic}.
In addition, the design of entropy models determine the bit rate needed to losslessly code the compressed representation.
Even with the identical analysis and synthesis transformations, the LIC with entropy model that better captures image statistic can achieve better rate-distortion efficiency, as has been shown in~\cite{minnen2018joint}.

\section{Conclusion}
\label{sec:conclusion}
In this paper, we analyze LICs from the perspective of basis decomposition.
By showing the basis functions of LICs, we empirically conclude that LIC transformations can be interpreted as orthogonal transformations in a non-linear fashion.
Our results provide better understanding of how LICs work and bring insights to the future development of learned image compression.


{
\bibliographystyle{IEEEtran.bst}
\bibliography{references.bib}
}

\end{document}